\documentclass[
reprint,
superscriptaddress,
 amsmath,
 amssymb,
 aps,
 longbibliography
]{revtex4-2}

\usepackage{cancel}
\usepackage{dcolumn}
\usepackage{bm}
\usepackage[usenames,dvipsnames]{color}
\usepackage{soul}
\usepackage{enumitem}
\usepackage{xspace}
\usepackage{booktabs}
\usepackage{lipsum}
\usepackage[ISO]{diffcoeff}
\usepackage{ragged2e}
\usepackage{mathrsfs}
\usepackage[normalem]{ulem}
\usepackage{wrapfig}
\usepackage{esvect} 
\usepackage{titletoc}
\usepackage{chngcntr}
\usepackage{fontawesome5}
\usepackage{mleftright}
\usepackage[nodayofweek]{datetime}
\usepackage{comment}

\usepackage{graphicx}
\usepackage{array}
\usepackage{siunitx}
\usepackage{bookmark}
\usepackage{gensymb}
\usepackage{textgreek}
\usepackage[section]{placeins}
\usepackage{xcolor}
\usepackage{dblfloatfix}

\definecolor{firebrick}{HTML}{B22222}

\hypersetup{
    colorlinks=true,       
    linkcolor=firebrick,          
    citecolor=firebrick,        
    filecolor=firebrick,      
    urlcolor=firebrick           
}

\definecolor{orcid-green}{RGB} {166, 206, 57}
\newcommand{\MYhref}[3][blue]{\href{#2}{\color{#1}{#3}}}%

\newcommand{\red}{\textcolor{red}}

\newcommand{\DU}[1]{{\textcolor{orange}{{[\bf DU: #1]}}}}

\newcommand{\kB}{k_\text{B}} 

\begin{document}

\newcommand{\titleinfo}{\boldmath Picometer control of a levitating milligram gravity sensor}
\newcommand{\authorinfo}{Dennis G.~Uitenbroek, Jurriaan Langendorff, and Tjerk H.~Oosterkamp}

\title{\titleinfo}

\author{Dennis G.~Uitenbroek\,\MYhref[orcid-green]{https://orcid.org/0009-0008-3425-6406}{\faOrcid}}
\affiliation{Leiden Institute of Physics, Leiden University, P.O. Box 9504, 2300 RA Leiden, The Netherlands.}

\author{Jurriaan Langendorff\,\MYhref[orcid-green]{https://orcid.org/}{\faOrcid}}
\affiliation{Leiden Institute of Physics, Leiden University, P.O. Box 9504, 2300 RA Leiden, The Netherlands.}

\author{Tjerk H.~Oosterkamp\,\MYhref[orcid-green]{https://orcid.org/0000-0001-6855-5190}{\faOrcid}}
\affiliation{Leiden Institute of Physics, Leiden University, P.O. Box 9504, 2300 RA Leiden, The Netherlands.}

\begin{abstract}
\noindent
Due to their exceptional isolation from the environment, magnetically levitated particles are explored as extremely sensitive mechanical sensors. For future gravity experiments on quantum superpositions, such systems need to be cooled close to their ground state. To demonstrate the combination of state of the art vibration isolation, milligram levitated high Q mechanical resonators and position detection with low noise, we present linear feedback cooling of a magnetically levitated gravity sensor to below 2 picometer amplitude and below 10 millikelvin mode temperature for two translational modes (the $x$- and $y$-mode) simultaneously. The sensor is a levitating permanent magnet in a type I superconducting trap, where its six resonance frequencies are measured with a superconducting coil coupled to a DC SQUID.  This signal is measured with a lock-in amplifier and a feedback signal is sent to a piezoelectric actuator, allowing the cooling of resonant modes  at \SI{50.6}{} and \SI{68.0}{Hz} simultaneously. These two translational modes have Q factors of \SI{3.8e6}{} and \SI{5.5e6}{} respectively. The experiment is mounted inside a dry dilution refrigerator where it is vibrationally attenuated with \SIrange{110}{130}{dB} at these frequencies. In this work, we discuss future improvements on the setup which may enable quantum ground state cooling on a magnetically levitated particle, that has previously been shown to be a gravitational sensor.
\end{abstract}
\maketitle


\section{Introduction}\label{sec:introduction}

Large mass quantum experiments are the key ingredient in closing the gap between microscopic quantum systems and macroscopic general relativity. By pushing the quantum effects to larger and heavier scales, an experiment combining these two theories could be enabled. Observing quantum mechanical effects in a mechanical system requires a crucial first step: cooling a particle close to the mechanical quantum ground state. This has been a goal within the physics community on various platforms using different cooling methods. Additionally, one would like to achieve this quantum regime in a mass regime where gravity becomes relevant.

In early work, micromechanical membranes were shown to be cooled to the quantum ground state~\cite{Connell2010, Teufel2011, Chan2011, Safavi2012, Jockel2015}. These type of systems will inevitably suffer from clamping losses, due to the suspended nature of these experiments.

In an effort to avoid clamping losses, levitated systems can be studied. Trapped ions were the first to be cooled to the mechanical groundstate, using laser sideband cooling~\cite{Diedrich1989, Monroe1995, Leibfried2003}. Although impressive results are accomplished in this field, these type of experiments are limited to very few atoms in scale.

In optical levitation, groundstate cooling has been achieved~\cite{Delic2020, Tebbenjohanns2021, Magrini2021, piotrowski2023} since a couple of years. These experiments are performed at relatively high frequencies (\SI{10}{k\hertz} - \SI{1}{M\hertz}) compared to magnetic levitation. At these frequencies mechanical vibrations are less present and can be readily attenuated. This is a very promising platform for pushing the limits of quantum mechanical effects. However, these particles have a size around \SI{100}{nm} and a mass in the order of femtograms. This is already a large step in macroscopicity compared to ions, but still far from a macroscopic particle that would generate a measurable amount of gravity.

A different levitation platform, magnetic levitation, makes use of magnetic forces. Two main advantages of this platform are, no limitation in levitated mass due to the magnetic forces involved, and no internal heating of the particle as in the above-mentioned levitation systems. With the larger mass the frequencies will decrease (compared to optical levitation). At these lower frequencies dampening the mechanical vibrations is more challenging.

Different types of magnetic levitation exist, we will only discuss superconducting levitation here, superconducting levitation can be divided in active~\cite{Bob, Gutierrez2020, Gutierrez2022, Hofer2023, Gutierrez2023, Schmidt2024, Paradkar2025, Hansen2026} and passive systems~\cite{Timberlake2019, Vinante2020, Vinante2022, Timberlake2024}. Active systems send a current through coils that levitate a piece of superconducting material by using the perfect diamagnetic properties of the superconductor. The passive systems use the same perfect diamagnetic property due to the Meissner effect and let the superconductor expel the magnetic field, coming from a permanent magnet. Both mechanisms are, in principle, scalable to sizes greater than \SI{1}{mm}. The advantage of a passive system is the absence of noise sources from the active components, e.g. the current noise in the coils.

Passive magnetic levitation with a submillimeter particle of \SI{0.43}{mg} has already been demonstrated in a gravitational experiment~\cite{Gravity}. In this article, a version of this setup with state-of-the-art vibration isolation, a massive levitated particle with high Q factors, and low noise readout are all together combined to reduce the motion of the particle to below two picometer of RMS amplitude. This achieved motion is a benchmark for the performance of the system, this can only be achieved if all necessary experimental conditions are fulfilled simultaneously. Performing ground state cooling on these type of systems is possible in theory, but experimentally very challenging. The zero point motion scales with $\sim 1/\sqrt{m \omega}$. In magnetic levitation the relatively large masses in combination with low frequencies result in a small zero point motion, \SI{0.7}{fm} for the setup used here.


In this experiment a phonon number in the order of \SI{e6}{} is reached. The theoretical minimum phonon limit with current detection and force noise is in the order of \SI{e5}{} phonons. With improvements on vibration isolation, Q factors, detection coupling and SQUID noise, we think it would be possible to cool this system to the ground state. 
Since this setup has already been used for gravitational measurements~\cite{Gravity}, combining this system with the feedback cooling presented in this paper, this experiment demonstrates to be a promising platform towards combining both gravitational interaction and quantum mechanical effects in a single experiment.


\section{Experimental setup}\label{sec:setup}
\begin{figure}[b]
    \centering
    \includegraphics[width=\columnwidth]{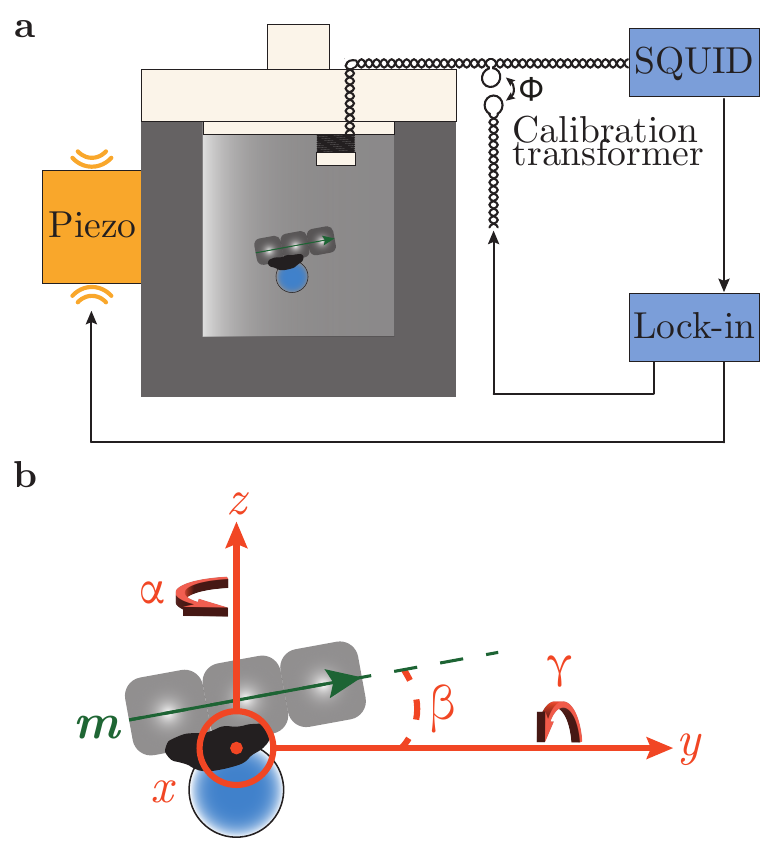}
    \caption{\textbf{(a)} A schematic outline of the experimental setup, the levitating permanent magnet is shown inside the superconducting trap. A pick-up coil attached to the lid of the trap picks up the varying flux of the moving levitated magnet and is detected using a two stage SQUID amplifier. We feedback cool the magnets motion by applying a phase shifted tone shaking the trap housing through a piezo and lock-in amplifier. \textbf{(b)} The defined axes of the levitated magnet.}
    \label{fig:setup-schematic}
\end{figure}

The experiment consists of a permanent magnet levitating in a superconducting trap by Meissner repulsion. The motion of the magnet is measured using a Superconducting QUantum Interference Device (SQUID) measured by a lock-in amplifier. The lock-in then calculates a feedback signal and sends this to a piezoelectric actuator, schematically shown in Fig.~\ref{fig:setup-schematic}a.

\begin{figure}[h]
    \centering
    \includegraphics[width=\columnwidth]{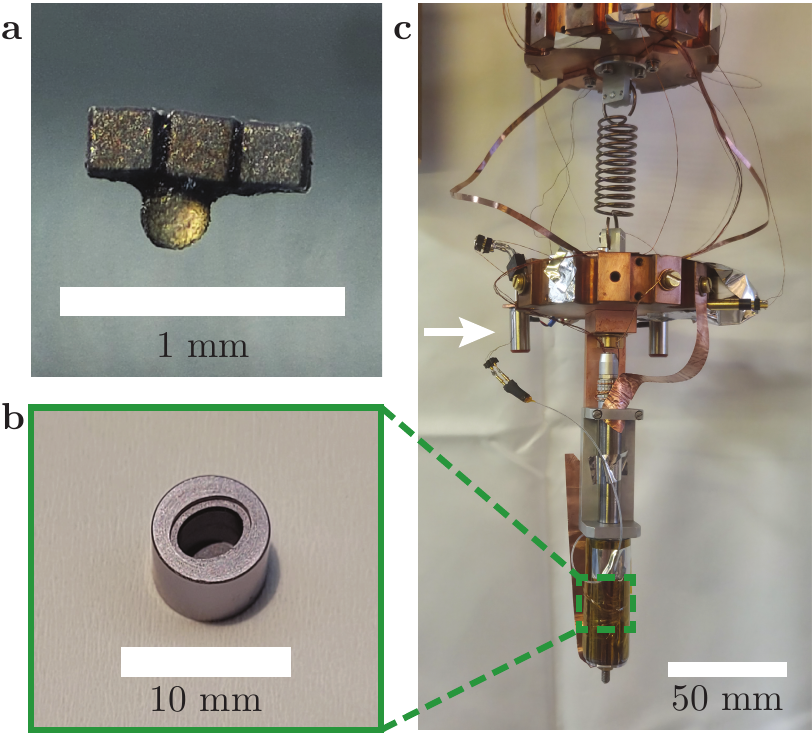}
    \caption{Overview of the experimental setup. \textbf{(a)} A close-up of the permanent magnet used in the experiment. It is composed of three \SI{0.25}{mm} cubic neodymium iron boron parylene coated magnets stacked and glued together. A glass sphere with a diameter of \SI{0.20}{mm} is attached to this stack of magnets using Stycast, to break rotational symmetries. \textbf{(b)} Picture of the elliptically shaped superconducting trap, made from tantalum. \textbf{(c)} Closeup of the bottom part of the multi-stage mass spring system where the magnet, trap and piezo are located. The trap is located inside the magnetic shielding, indicated with the green dashed lines. The piezo is attached to the lowest copper mass and indicated with a white arrow.}
    \label{fig:setup-pictures}
\end{figure}

The permanent levitating magnet is composed of three \SI{0.25}{mm} by \SI{0.25}{mm} by \SI{0.25}{mm} cubic neodymium iron boron parylene coated magnets stacked and glued together, using Stycast. A glass sphere with a diameter of \SI{0.20}{mm} is attached to this stack of magnets, again using Stycast, see Fig.~\ref{fig:setup-pictures}a. The glass sphere is placed off center to break symmetry. Typical permanent remnant magnetisation of these magnets is \SI{1.4}{T}, although it is believed that at low temperatures this is reduced by \SIrange{20}{25}{\%}~\cite{NdFeB_SmCo_Technotes}. The estimated mass of this composed magnet and glass sphere is \SI{0.38}{mg}.

The magnet levitates in a type I superconducting trap made of tantalum, shown in Fig.~\ref{fig:setup-pictures}b. This trap has an elliptical shape, \SI{3.0}{mm} by \SI{4.0}{mm} as the minor and major axes. The pick-up coil is placed at a height of \SI{4.0}{mm} from the bottom of the trap. The calculated levitation height of the magnet is \SI{1.96}{mm}~\cite{Vinante2020}. The pick-up coil is able to detect the motion of the magnet, via the induced current in the pick-up coil. The pick-up coil is part of the detection scheme consisting of the pick-up coil, the calibration transformer, and the SQUID input coil. The calibration transformer is used to calibrate the coupling between the detection circuit and the six different degrees of motion of the magnet. A schematic overview of the detection can be seen in Fig.~\ref{fig:setup-schematic}a.


\clearpage
\begin{widetext}
\begin{figure}[t]
    \includegraphics[width=\textwidth]{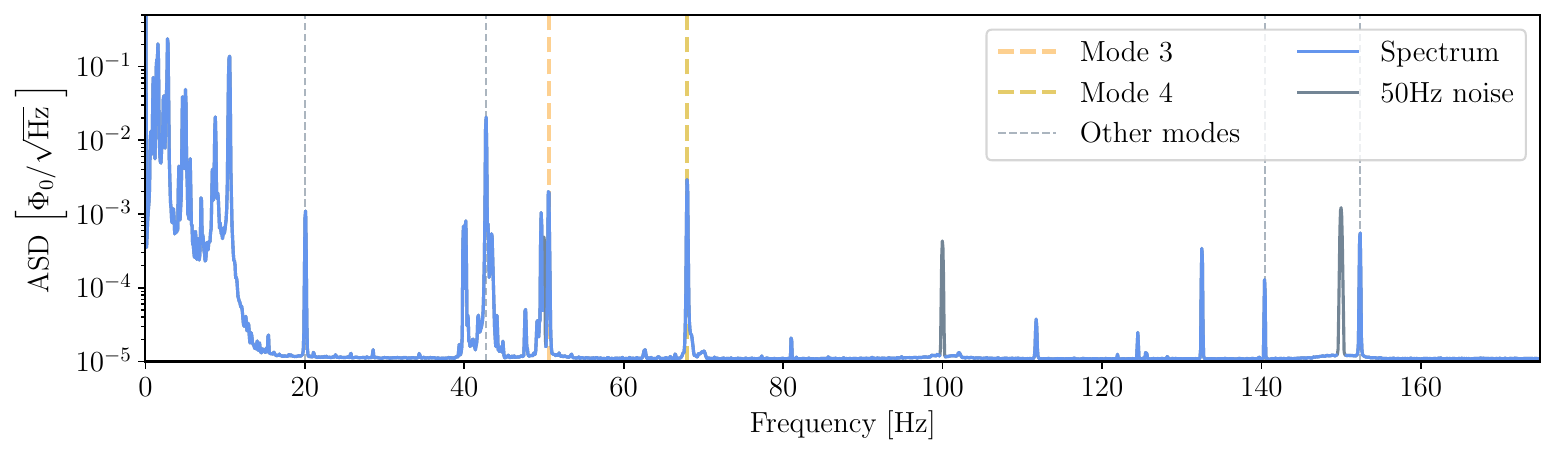}
    \begin{minipage}{\textwidth}
    \caption{Typical amplitude spectral density (ASD) of the levitated magnet, the resonant modes of the magnet are indicated with the dotted lines. In this figure the \SI{50}{Hz} European electrical noise is grayed out.}
    \label{fig:spectrum} \end{minipage}
\end{figure}
\end{widetext}

Vibrational noise is suppressed by a multistage mass spring system both in the vertical and lateral directions. The lowest two masses and one of the springs can be seen in Fig.~\ref{fig:setup-pictures}c. The experiment is rigidly attached to a copper mass, which is suspended from two other copper masses with springs in between. The lowest resonance frequencies are at \SI{0.6}{Hz} and \SI{1}{Hz} along the lateral and vertical directions, respectively. These three stages are thermalised to the millikelvin stage inside a dry dilution refrigerator. In the current configuration, the mass spring system is suspended from the still plate in order to utilize the additional available vertical length. For further attenuation, the still plate is suspended by springs from the \SI{3}{K} plate. 

The cryostat as a whole is rigidly attached to a 25-metric ton concrete block located in the cellar of the building, placed on pneumatic dampers to limit vibrations. The pulse tube and circulation pumps are rigidly attached to the building via a second frame. 



The SQUID signal is processed at different frequencies simultaneously by a lock-in amplifier, which then provides a linear feedback signal which is sent to the piezo with the proper phase shift, to decrease the amplitude of the motion of the particle. The feedback gain can be altered to vary the effective damping.


\subsection{Mode identification}
Once the magnet is levitating, many clear peaks in the spectrum are visible, shown in Fig.~\ref{fig:spectrum}. Six peaks correspond to six degrees of freedom, verified by a magnetic drive, of the levitating magnet, three translational and three rotational. Currently, for our system, assigning mode frequencies to their respective modes of motion is not possible with complete confidence. Since we have only one detection coil, there is no distinction in the measurement process for different degrees of freedom. Although the magnetic flux couples differently to different modes, we cannot yet use this information to assign which frequency correspond to which mode because we cannot assume that the equilibrium position of the magnet is exactly in the center of the trap.

We find different resonance frequencies between cooldowns, even when we don't make any changes to the setup. Frequencies change when warming up the trap above the critical temperature of the superconductor and immediately cooling down again. We attribute this to some flux from the magnet remaining trapped in the superconductor as the superconductor is cooled through its transition from being a normal metal to becoming a superconductor, although the type I superconductor expels most of the magnetic field. We have calculated the expected frequencies for the modes with analytical models~\cite{Vinante2022,Francis2025}.

The order, from low frequency to high frequency, we find from the analytical calculations is always: $\gamma$, $z$, $y$, $x$, $\beta$, $\alpha$ with the axes shown in Fig.~\ref{fig:setup-schematic}b. Also when varying the dimensions of the trap slightly around the values of our particular trap, the values for the frequencies change, but the order of the modes does not change. This means, although not with complete confidence, that the described mode 3 would belong to the $y$-mode and mode 4 would belong to the $x$-mode of the particle.

We also created an offset in the inclination angle ($\beta$) of the particle with respect to the horizontal axis. This is done by purposely gluing the glass bead asymmetrically below the magnet, to avoid that one of the modes would couple very weakly to the pick-up coil. However, we don't know at exactly what inclination angle the magnet will levitate in the trap. This also does result in the same order of frequencies.


Based on coupling and sensitivity, mode 3 and mode 4 were chosen to cool using linear feedback with the piezo. 

\subsection{Calibration}
The motion of the particle is calibrated using the calibration transformer, schematically shown in Fig.~\ref{fig:setup-schematic}a. A current is sent through this coil using the lock-in amplifier, the energy sent into the superconducting detection circuit and the energy gain in the motion of the particle are both measured by SQUID detection. These measurements together with known properties of the circuit and the magnet, provides the absolute calibration of the motion, in units $\SI{}{V/m}$. This method is explained in Supp. Mat. C in Fuchs et al.~\cite{Gravity}. The main result of this derivation is equation S12 in this publication, giving the relationship between the measured voltage in the SQUID based measurement and the motional amplitude of the magnet, reading
\begin{equation}
    \frac{dV}{dx} = 
    \frac{dV}{d\Phi_\text{SQ}}\cdot\frac{d\Phi_\text{SQ}}{dI}\cdot\frac{dI}{d\Phi}\cdot\frac{d\Phi}{dx} =
    \frac{dV}{d\Phi_\text{SQ}} \cdot \frac{M_\text{in,SQ}}{L_\text{total}}\cdot \frac{d\Phi}{dx}.
\end{equation}

Calculating these values for the two cooled modes results in $\frac{dV}{dx} = \SI{4.7}{{\frac{V}{\micro m}}}$ for mode 3 and $\frac{dV}{dx} = \SI{3.0}{{\frac{V}{\micro m}}}$ for mode 4, with a relative error of 8.5\%, which is dominated by the error in the mass.

For a rotational mode the same method can be applied, and a value in units of \SI{}{V/rad} can be derived. To calculate a mode temperature this would result in a 
time average over the angle squared $ \left< \frac{1}{2} \kappa \theta^2 \right>_t$ in analogy to $ \left< \frac{1}{2} k x^2 \right>_t$, where $\kappa$ is the rotational stiffness, which calculates the energy in this rotational mode. If such a rotational mode is treated as a translational mode by accident and the mode temperature is calculated, the same result would be obtained as in the treatment of a rotational mode. 

Note that other peaks in the spectrum are due to magnetic or mechanical noise, but these cannot be excited with the calibration coil.

\subsection{Cooling}
The response of the levitating magnet to the linear feedback signal can be calculated from the following equation of motion
\begin{equation}
    m \ddot{x}(t) + \gamma_\text{0} \dot{x}(t) + kx(t) = F_\text{tot}(t) = F_\text{th}(t) + F_\text{FB}(t).
\end{equation}

In this equation $m$ is the mass of the magnet, $x(t)$ is the position, $\gamma_0$ is the mechanical damping in the system, with units $\SI{}{N/(m/sec)} = \SI{}{kg/sec}$. Furthermore, $k$ is the spring constant of the mode, $F_\text{tot}$ is the total force acting on the particle, $F_\text{th}$ is the random thermal force and $F_\text{FB}$ is the feedback force. This feedback force provides a feedback damping ($\Gamma_\text{FB}$) for which the proportional gain, $g$, can be set, this is an additional damping to the damping in the system ($\Gamma_\text{0}=\gamma_0/m$). The total damping then becomes $\Gamma_\text{tot}=\Gamma_\text{0}+\Gamma_\text{FB}=(1+g)\Gamma_\text{0}$~\cite{Kleckner2006}. When the phase of the feedback force is chosen correctly such that the feedback force is proportional to the velocity of the particle, this equation of motion has a corresponding power spectral density (PSD) with units $\SI{}{m^2/Hz}$

\begin{equation} \label{eq:Sxx}
    S_{xx}(\omega) = \frac{4 \kB T \Gamma_\text{0}}{m} \frac{1}{\left( \omega_0^2 - \omega^2 \right)^2 + \omega^2 \left(\Gamma_\text{0} + \Gamma_\text{FB} \right)^2}.
\end{equation}

For center of mass modes the thermal force noise is given by $S_{FF} = 4 \kB T m \omega_0 / Q$ for rotational modes the torque noise equals $S_{\tau \tau} = 4 \kB T I \omega_0 / Q$. Here, $\kB$ is the Boltzmann constant, $T$ is the effective temperature formed by the thermodynamic temperature and vibrations at the resonance frequencies, $m$ is the mass, $\omega_0$ is the resonance frequency, $Q$ is the quality factor and $I$ is the moment of inertia. 

This power spectral density can also be expressed as an amplitude spectral density (ASD) in units of $\SI{}{m/\sqrt{Hz}}$ by taking the square root. Because the quality factor is very high, it is sufficient to integrate over a bandwidth ($\text{BW}$) $\left[ f_0-3f_0/Q, f_0+3f_0/Q \right]$ to get 99 percent of the energy. Integrating over this bandwidth, the amplitude ($A_\text{RMS}$) and energy in the mode are calculated. Next, using the equipartition theorem the mode temperature ($T_\text{mode}$) is determined

\begin{align} \label{eq:Tmode}
    T_\text{mode} = \frac{k}{\kB} \int\displaylimits_\text{BW} S_{xx}(\omega) d\omega .
\end{align}

\section{Results}\label{sec:results}
The described method has been applied to two translational modes simultaneously. The applied feedback signal is a phase shifted version of the measured signal, with a variable gain. The use of the lock-in amplifier ensures that we send a narrow band version of the measured signal, which is difficult with analogue band pass filters. The measured lock-in signal is Fourier transformed using the Welch method, to average over the measurement time. Finally, the bandpass filter of the lock-in is compensated numerically.

\begin{figure}[h!]
    \centering
    \includegraphics[width=\columnwidth]{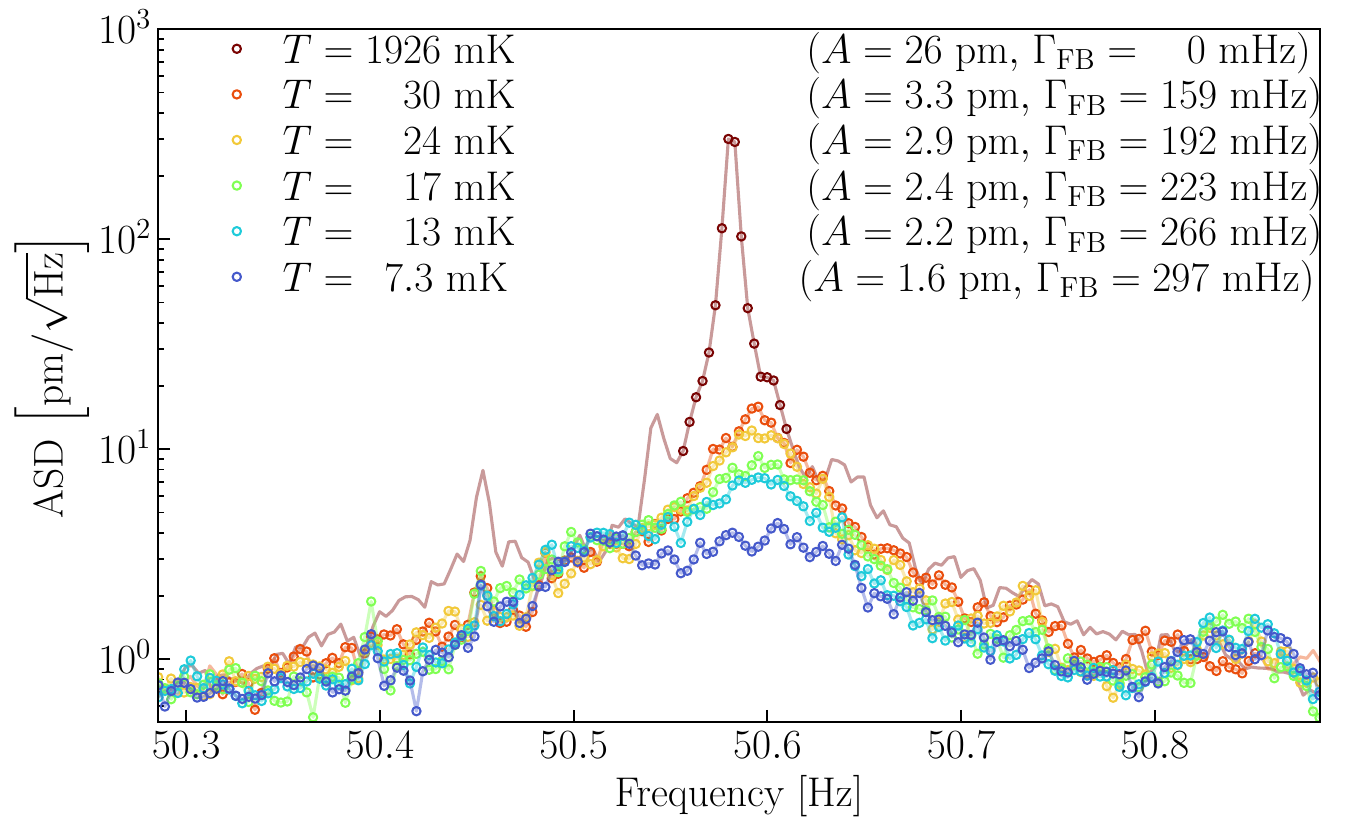}
    \caption{The linear feedback cooling on mode 3 is plotted. The different colored lines represent spectra taken with different linear gains, i.e. feedback dampings. The legend shows the achieved mode temperatures with corresponding RMS amplitude and feedback damping values. Equal colors in this plot and in Fig.~\ref{fig:cooling4} indicate measurements are taken simultaneously.}
    \label{fig:cooling3}
\end{figure}

\begin{figure}[h]
    \centering
    \includegraphics[width=\columnwidth]{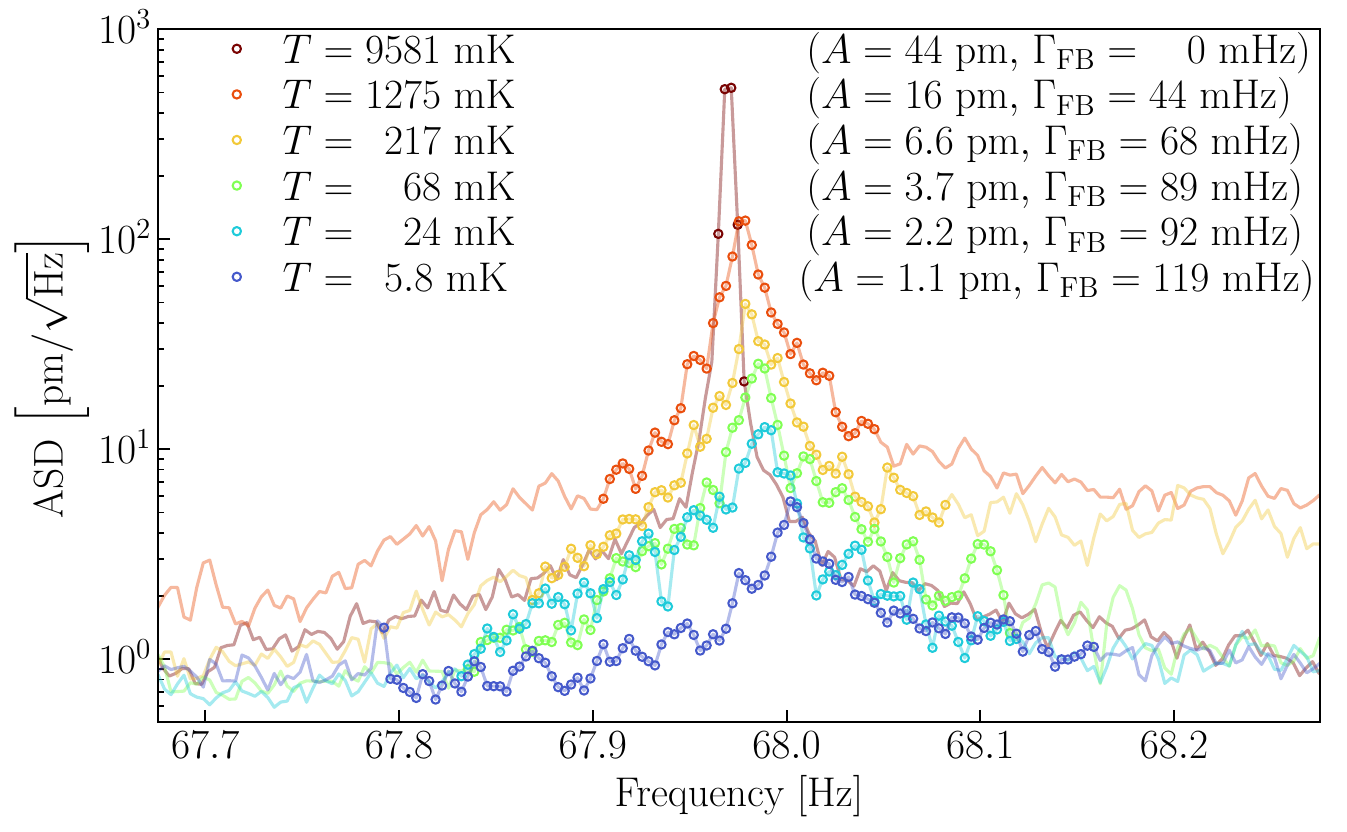}
    \caption{Similar to Fig.~\ref{fig:cooling3}, the linear feedback cooling on mode 4 is plotted. The different colored lines represent spectra taken with different linear gains, i.e. feedback dampings. The legend shows the achieved mode temperatures with corresponding RMS amplitude and feedback damping values. Equal colors in this plot and in Fig.~\ref{fig:cooling3} indicate measurements are taken simultaneously.}
    \label{fig:cooling4}
\end{figure}

The results for mode 3 and mode 4 are plotted in separate figures. In the legend, the values of the achieved mode temperatures with corresponding RMS amplitude and proportional gain are given. Increasing the proportional gain results in lowering the peak, leading to an effective broadening of the peak, as expected. Same colors in the legends of Fig.~\ref{fig:cooling3} and Fig.~\ref{fig:cooling4} correspond to the same moment in time. 

Mode 3 has a resonance frequency of \SI{50.59}{Hz} and a Q factor of \SI{3.8e6}{}. The lowest observed mode temperature for mode 3 is \SI{7.3}{mK}, obtained with a feedback damping of $\Gamma_\text{FB}=\SI{297}{mHz}$ and has a corresponding RMS amplitude of \SI{1.6}{pm}. 

Mode 4 has a resonance frequency of \SI{67.98}{Hz} and a Q factor of \SI{5.5e6}{}. The same cooling mechanism is applied to mode 4, in this case the lowest observed mode temperature is \SI{5.8}{mK} simultaneously with the \SI{7.3}{mK} of mode 3. Here, a feedback damping of $\Gamma_\text{FB}=\SI{119}{mHz}$ is applied and a corresponding RMS amplitude of \SI{1.1}{pm} is measured.

Both plots show lines and dots, the lines represent the measured data. The part of the data that falls within the mentioned bandwidth ($\text{BW}$) of $\left[ f_0-3f_0/Q, f_0+3f_0/Q \right]$ is visualised with the dots in both Fig.~\ref{fig:cooling3} and Fig.~\ref{fig:cooling4}. The shown RMS amplitude and mode temperature are therefore based on the cumulative sum of these dots. 

These results are achieved while not only doing active feedback on mode 3 and 4, but also on some of the mass spring system modes and on mode 2 with a second piezo. Active feedback is applied on these frequencies continuously to ensure a relatively quiet environment for the linear feedback on mode 3 and 4. During these measurements upconversion of noise peaks surrounding the resonance peaks is visible in the spectrum, as explained below.


Combining equations~\ref{eq:Tmode} and~\ref{eq:Sxx} the mode temperature and total damping are related as $T_\text{mode} \propto 1/\Gamma_\text{tot}$~\cite{Kleckner2006}. This is the case because the peak in the transferfunction goes down with increasing $g$ as $\frac{1}{(1+g)^2}$ and the bandwidth used in summation goes up as $\text{BW} \propto 1+g$. Together this causes the energy to be proportional to $\frac{1}{1+g}$. 

\begin{figure}[h]
    \centering
    \includegraphics[width=\columnwidth]{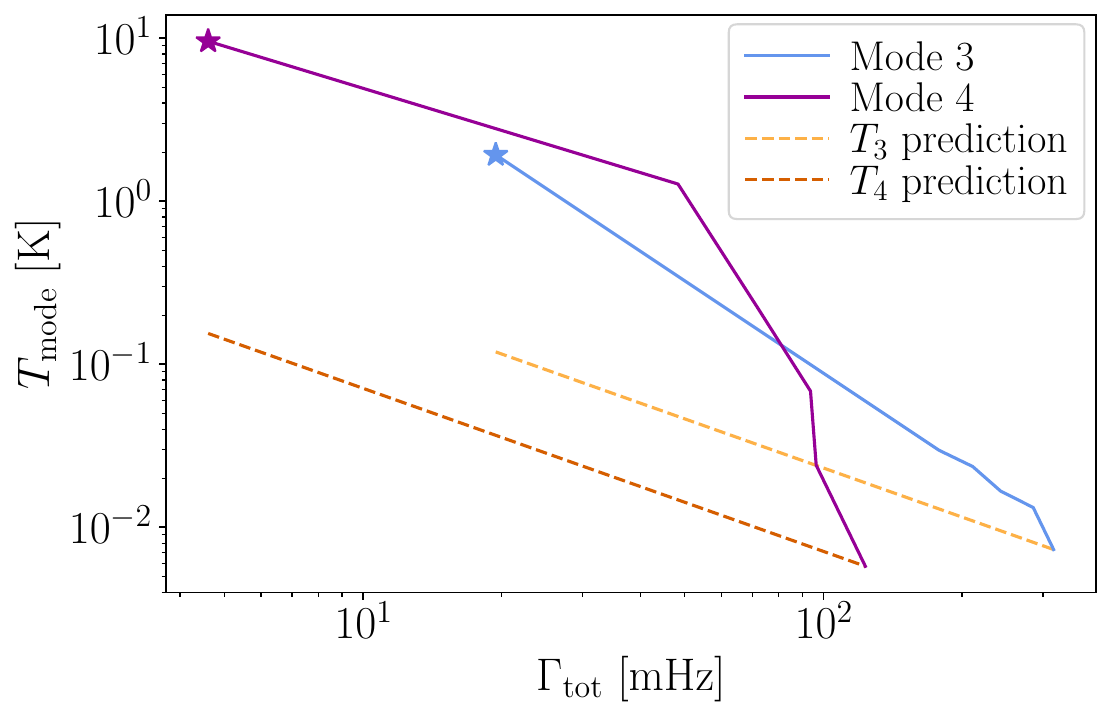}
    \caption{Measured mode temperature of both mode 3 and 4 plotted as a function of the total damping $\Gamma_\text{tot}$. The dashed lines are extrapolations from the lowest achieved mode temperatures, back to the mode temperatures that one would expect if there were no non-linear mixing. This is done using the expected $1/\Gamma_\text{tot}$ scaling. For both modes the data point without feedback damping ($\Gamma_\text{0}$) is labeled with a star.}
    \label{fig:modeTdamping}
\end{figure}

During the cooling experiments for each proportional gain we used, the ringdown time of the mode was measured, and thus the damping. These are used to plot the measured mode temperature against the total damping, visible in Fig.~\ref{fig:modeTdamping}. In the figure, not only the measured mode temperatures are visible (solid lines), but also the predicted mode temperatures (dashed lindes) of both modes are shown. The predictions are extrapolations from the lowest achieved mode temperatures back to zero feedback gain $g$, with the mentioned $1/\Gamma_\text{tot}$ scaling. The measured data point without feedback damping is labeled with a star, these values are $\Gamma_\text{0} = \SI{19}{mHz}$ for mode 3 and $\Gamma_\text{0} = \SI{4.6}{mHz}$ for mode 4.

In the figure, it is clear that the measured mode temperatures for both modes are decreasing faster than the prediction when increasing the damping. This is attributed to amplitude dependent mode mixing due to nonlinearities of the magnetic trap. At larger amplitudes, low frequency noise is coupled relatively more strongly into the amplitude of the cooled modes. We discuss this in more detail in the following section, where we illustrate the mode mixing in combination with feedback cooling with Fig.~\ref{fig:upconversion_mode_3}.

\subsection{Mode mixing due to nonlinearities}
\begin{figure}[h]
    \centering
    \includegraphics[width=\columnwidth]{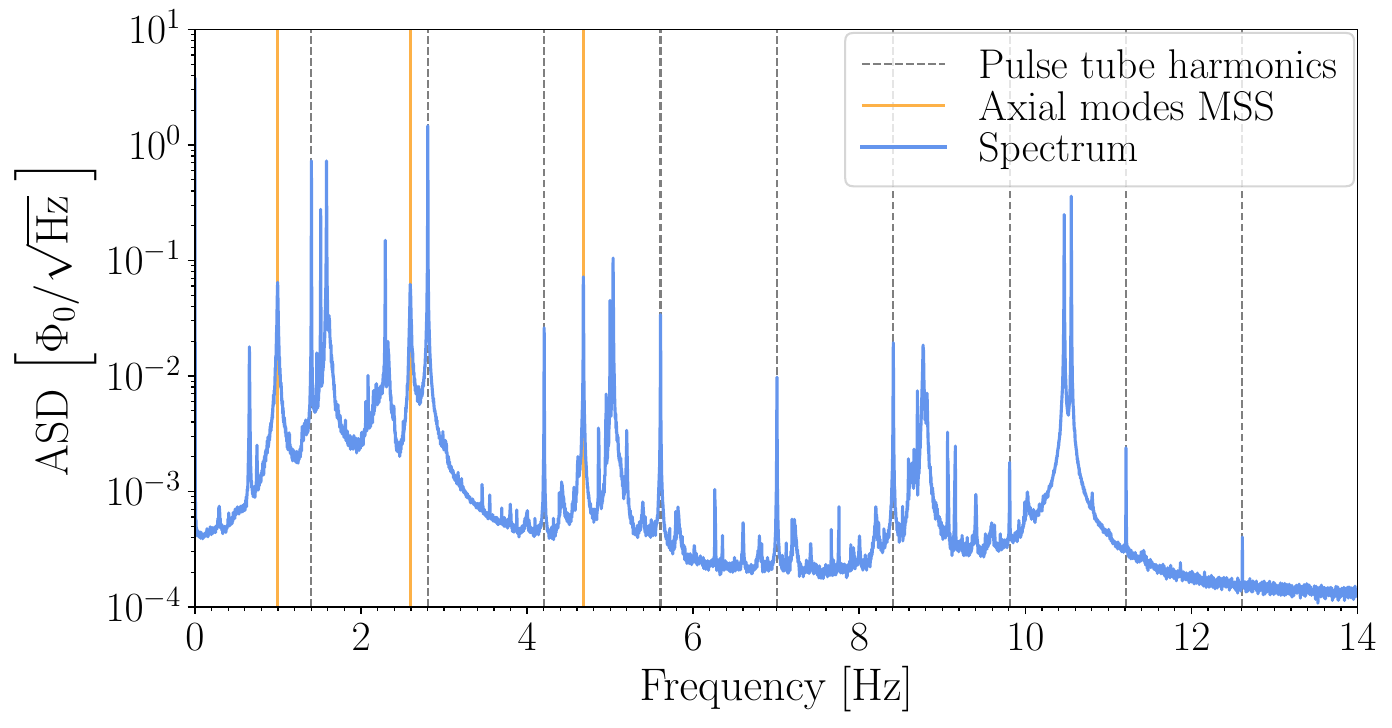}
    \caption{Typical ASD of the resonances of the mass spring system as measured by the SQUID that measures the relative motion between the levitated magnet and pick-up coil. Multiples of the pulse tube frequency (\SI{1.401}{Hz}) are indicated with the vertical dashed lines. The axial modes of the mass spring system (MSS) as simulated are indicated with the vertical lines.}
    \label{fig:spectrum_mass_spring} 
\end{figure}

The signal detected by the SQUID contains not only the six modes related to the six degrees of motion of the magnetic particle (three librations and three translations). It also contains other vibrational movement of the setup. Modes of the mass spring system are most prominent in the \SIrange{0.5}{15}{Hz} range, see Fig.~\ref{fig:spectrum_mass_spring}. 

Many of these peaks in the low frequency spectrum can be attributed to vibrations caused by either the cooling equipment or the resonances of the vibration isolation system. Using a finite element model, we can link certain frequencies to modes of the mass spring system (MSS).

\begin{figure}[h]
    \centering
    \includegraphics[width=\columnwidth]{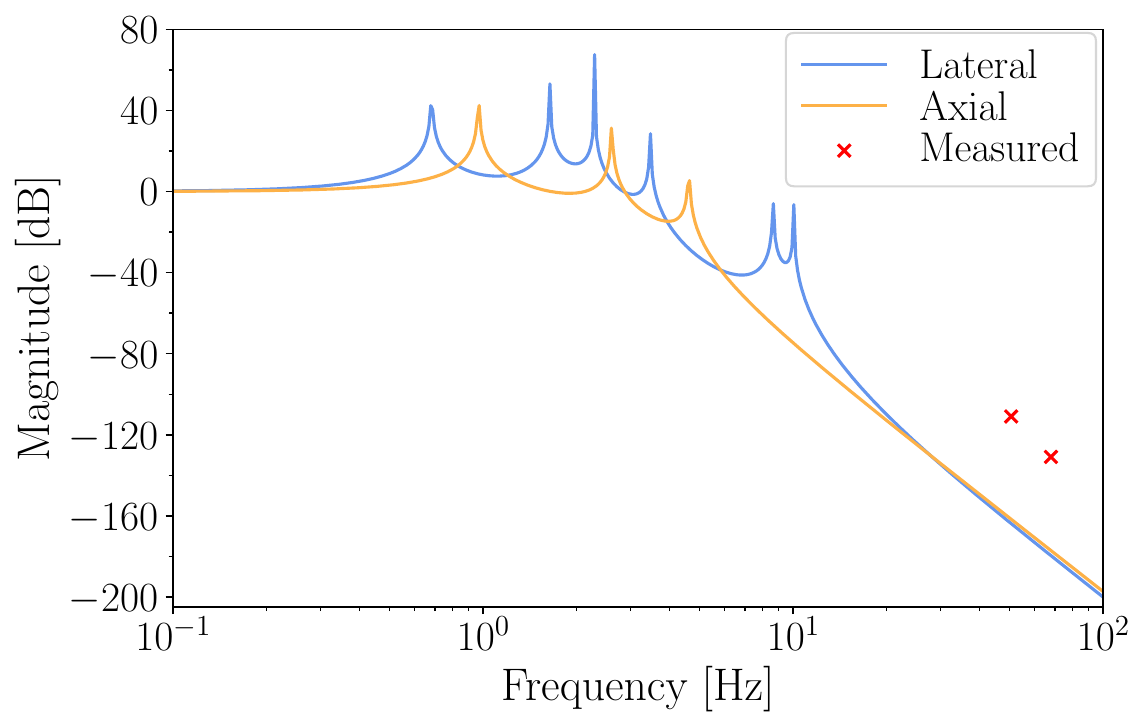}
    \caption{Bode magnitude plot of the attenuation of both the lateral and axial motion of the mass spring system (MSS), simulated with a finite element model. The measured attanuation on modes 3 and 4 are shown as red crosses.}
    \label{fig:spectrum_attenuation}
\end{figure}

In Fig.~\ref{fig:spectrum_attenuation}, a finite element simulation transfer function is plotted of both the lateral and axial motion of the MSS to the experiment. To generate these plots, a simplified model is solved using a finite element method. Here, the three masses are simulated as perfect cylinders, the helical springs are idealized as perfectly elastic snare springs, and their spring constants are physically measured at room temperature and extrapolated to cryogenic temperatures.

We have measured \SI{111}{dB} attenuation on mode 3 and \SI{131}{dB} attenuation on mode 4 in energy, in agreement with typical values with experimental runs from before for this system. This is consistent with a finite element simulation, showing between \SI{160}{} and \SI{180}{dB} under idealized conditions in the frequency range of \SIrange{50}{70}{Hz}; see Fig.~\ref{fig:spectrum_attenuation}. 
We do not think that external vibrations at the frequency of mode 3 and mode 4 are limiting the mode temperature. If this were the case, the position noise of the trap which holds the Zeppelin, would move by an amount equal to the force noise divided by the spring constant of the trap. Given that the force noise of mode 3 is approximately $\SI{60}{aN/\sqrt{Hz}}$, this would amount to a motion of approximately $\SI{1.5}{fm/\sqrt{Hz}}$.
Multiplying this position noise by the measured attenuation, yields a position noise at the room temperature part of the cryostat of $\SI{0.5}{nm/\sqrt{Hz}}$ or $\SI{150}{nm/s/\sqrt{Hz}}$.
For mode 4 these numbers are: force noise $\SI{125}{aN/\sqrt{Hz}}$, position noise of trap $\SI{1.8}{fm/\sqrt{Hz}}$, a position noise at the room temperature part of the cryostat of $\SI{5.7}{nm/\sqrt{Hz}}$ or $\SI{2.4}{\micro m/s/\sqrt{Hz}}$. 
This is two to three orders of magnitude above the vibrational signals we measure with a geophone at the top of the cryostat.

Instead of direct vibrational excitement, we believe that the upconversion of low frequency vibrations is causing the excitation of the magnet. Low vibrational noise couples in via nonlinearities in the magnetic potential. Furthermore, magnetic noise resulting from magnetic flux which escapes from the superconducting trap, limits the noise temperature to values higher than what might be expected from the thermodynamic temperature of the trap.  

We observe that the simulated axial modes (vertical motion) of the MSS align directly with some of the measured peaks in the spectrum. Other peaks in the spectrum can be explained by the lateral and rotational movement of the mass spring system. Comparing between the simulation and the measurements, the values do not all coincide, but some are shifted with the highest mode at \SI{10.5}{Hz}.

\begin{figure}[h]
    \centering
    \includegraphics[width=\columnwidth]{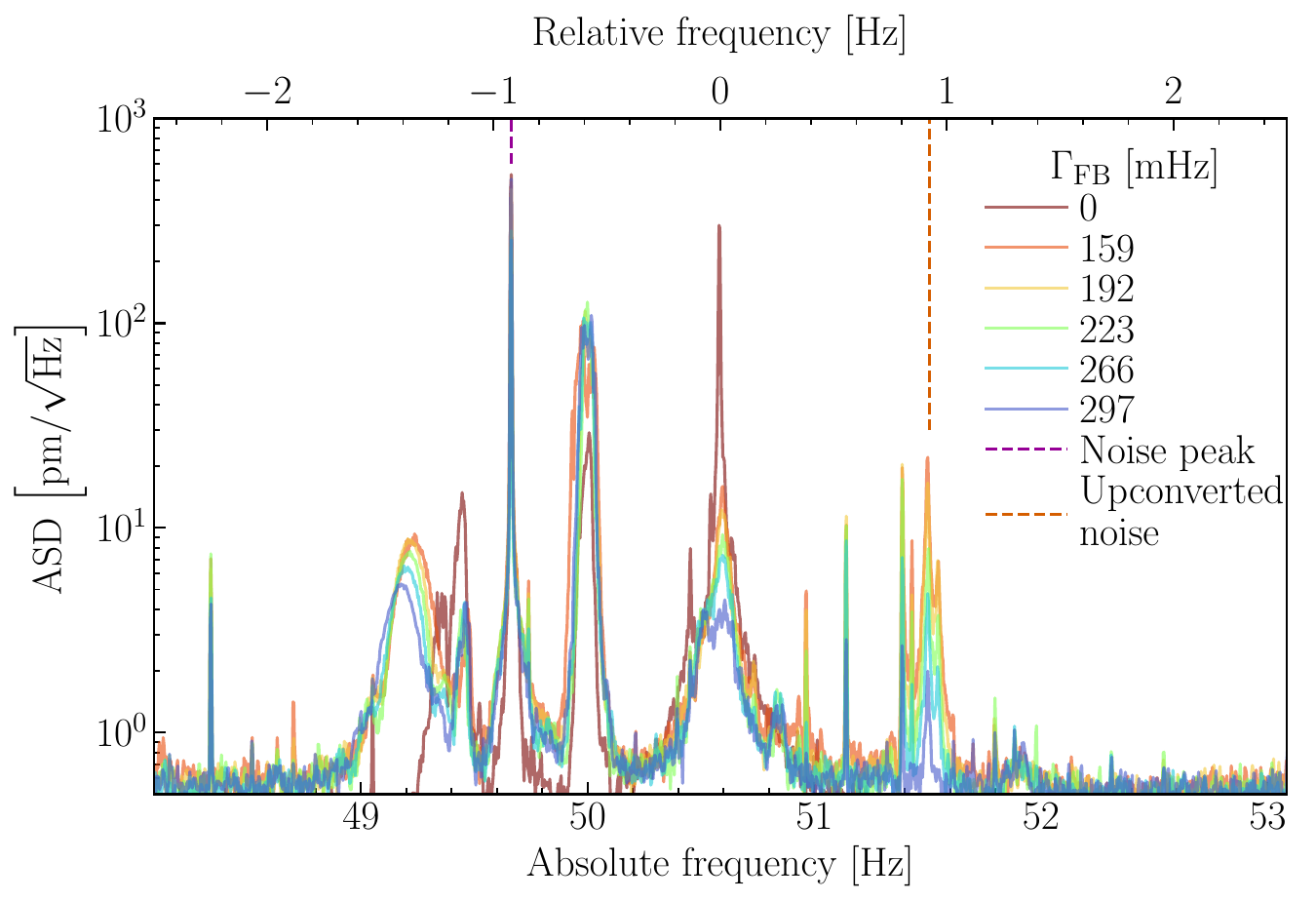}
    \caption{The spectrum around mode 3, where upconversion can be seen at for example \SI{51.51}{Hz}. The noise at \SI{49.67}{Hz} is reflected in the resonance and shows up as upconverted noise at \SI{51.51}{Hz}.}
    \label{fig:upconversion_mode_3} 
\end{figure}

During the experiment, we also observe peaks in the spectrum at frequencies that are the sum or difference of two peaks elsewhere in the spectrum. The most clear example can be seen in Fig.~\ref{fig:upconversion_mode_3} at \SI{51.51}{Hz}. This peak is caused by non linear mixing of the mode 3, which is at \SI{50.59}{Hz} with a strong excitation at \SI{49.67}{Hz}. This is a third order effect $(\SI{51.51}{Hz} = 2 \cdot \SI{50.59}{Hz} - \SI{49.67}{Hz})$. The upconverted noise peak at \SI{51.51}{Hz} is cooled simultaneously when mode 3 is cooled, indicating that this motional noise is coupled in through the motion of the magnet, the amplitude of the mode decreases with increasing gain and also the \SI{51.51}{Hz} noise decreases with increasing gain (but faster). 

We attribute the noise peak at \SI{49.67}{Hz} to a snare mode of the bottom spring of the vibration isolation system that we find in some finite element simulations.
We believe that all of our six modes can be excited by similar non-linear effects where multiple strong excitation peaks are mixed together and mixed with a continuum excitation to yield a frequency coinciding exactly with a mode frequency.

Please note that when we perform mode cooling and bring the particle motion of a translational mode down to a few picometers, the amplitude of the relative motion between the magnet and the trap at lower frequencies (\SIrange{0.5}{15}{Hz}) remains unchanged, and can be more than four orders of magnitude in amplitude larger as is visible in Fig~\ref{fig:spectrum}. For the \SI{50.6}{Hz} mode, mechanical vibrations of the compressor and circulation pumps of the dilution refrigerator (which are often slightly below \SI{50}{Hz}) mix with the mechanical modes of the vibration isolation system. We cannot account for all the peaks we see, but we attribute this to upconversion of several peaks mixing together to work against us. Also we cannot exclude that \SI{50}{Hz} electrical noise interferes with mechanical vibrations. Therefore we conclude that to improve the feedback cooling further we will need to also perform linear feedback cooling on the modes of the vibration isolation system.


\subsection{Phonon number}
After the modes have been cooled as described above, the mode temperature can be converted to an equivalent phonon number ($N_\text{ph}$) via:

\begin{equation}
    N_\text{ph} = \frac{\kB T_\text{mode}}{\hbar \omega_0}.
\end{equation}

In the case of mode 3 the lowest mode temperature corresponds to a value of $N_\text{ph, 3} = \SI{3.0e6}{}$ and for mode 4 a value of $N_\text{ph, 4} = \SI{1.8e6}{}$.

\section{Conclusion}\label{sec:conclusion}
In conclusion, we have performed linear feedback cooling to a magnetically levitated particle of submillimeter scale in a superconducting trap. Two different translational modes (the $y$- and $x$-mode) have simultaneously reached \SI{7.3}{mk} and \SI{5.8}{mK}. This is performed in a dry dilution refrigerator, with extensive vibration isolation. The mass of the levitated particle is \SI{0.38}{mg} and the feedback cooled phonon number is \SI{3.0e6}{} and \SI{1.8e6}{}, for mode 3 and 4 respectively.

Currently, we are limited by detection sensitivity and upconversion of resonances of the vibration isolation system below \SI{10}{Hz}. Improvements on the achieved mode temperatures in this experiment can be achieved with lower starting temperatures and/or lower force noise (i.e.  better vibration isolation), higher Q factors (which will also require an even better vibration isolation) and lower detection noise (which can be achieved by a larger coupling, i.e. placing the pick-up coil closer to the magnet). 

We can calculate the minimal achievable mode temperature and the corresponding phonon number of the current setup, due to force and detection noise:
\begin{equation}
    T_\text{min} = \frac{\omega_0}{2 \kB} \sqrt{ S_{FF} S_{xx_\text{det}}}
\end{equation}

Reading from both Fig.~\ref{fig:cooling3} and Fig.~\ref{fig:cooling4} a position noise of $\SI{1}{pm/\sqrt{Hz}}$, and using the uncooled ($\Gamma_\text{FB}=\SI{0}{mHz}$) mode temperatures for the force noise, for mode 3 this results in $T_\text{min, 3} = \SI{0.65}{mK}$ and $N_\text{ph, min, 3} = \SI{2.7e5}{}$, and for mode 4 this results in $T_\text{min, 4} = \SI{1.9}{mK}$ and $N_\text{ph, min, 4} = \SI{5.7e5}{}$ as the minimal achievable values with the current setup. 

Keeping the current detection sensitivity, but improving the vibration isolation such that upconversion is no longer an issue and the force noise is determined by the thermodynamic temperature of the environment (\SI{20}{mK}), the phonon numbers would fall to \SI{2.7e4}{} for mode 3 and \SI{2.6e4}{} for mode 4. The current magnet, at a frequency of \SI{50}{Hz}, has a zero point motion of \SI{0.7}{fm}, this corresponds to a mode temperature of \SI{2.4}{nK}.

In a more extreme case where the setup would be pushed towards its limits, with a thermodynamic temperature of \SI{2}{mK}~\cite{Loek, DeWitVibIso}, a 10 times higher Q factor, a detection sensitivity of $\SI{0.01}{pm/\sqrt{Hz}}$, we would be able to cool mode 3 to \SI{27}{} phonons and mode 4 to \SI{26}{} phonons.

Further improving the experiment and lowering the phonon amount towards the ground state, the back action noise from the SQUID will dominate over the thermal noise of the experiment, at some measurement coupling~\cite{ULDM}. This calls upon improving the SQUID noise to the quantum limit, since the SQUID we use has an energy resolution of approximately 10.000 times $\hbar$~\cite{ULDM, Ahrens2025, Vinante2021}.

Combining all these improvements, cooling and detection of the levitated particle in the quantum ground state should become possible.

\acknowledgments
We thank T.M. Fuchs, K. Heeck, B. Hensen, J. van Heijningen, M. Camp and E.J. Visser for useful discussions and experimental help. We thank M.L. Mattana and B. Hensen for their careful reading of the manuscript.
The authors acknowledge the support in this work by the EU Horizon Europe EIC Pathfinder project QuCoM (10032223). Furthermore, by the coordination of the Dutch Einstein Telescope collaboration, a joint effort of LIOF in cooperation with the Ministry of Economic Affairs and Climate Policy, the Ministry of Education, Science and Culture, and Nikhef.

\bibliography{main}






\end{document}